%% file: gce.tex
\documentclass[galley,usenatbib]{mnras}
\usepackage{ textcomp }

\newcommand{\sdssoneninef}{SDSS~J193712.01+502455.5~}
\newcommand{\sdsszerosixf}{SDSS~J064655.6+411620.5~}
\newcommand{\sdssonenine}{J1937+5024~}
\newcommand{\sdsszerosix}{J0646+4116~}

\include{journal_abbv}
\usepackage{graphicx}	
\usepackage{amsmath}	
\usepackage{amssymb}	
\usepackage{multicol}        
\usepackage{bm}		
\usepackage{pdflscape}	
\usepackage{float}
\usepackage{subcaption}
\usepackage{longtable}
\usepackage{journal}
\hbadness=\maxdimen
\vbadness=\maxdimen

\title[Likely GC Escapees in the Galactic Halo]{ A high-resolution spectroscopic study of two new Na- and Al-rich field giants - likely globular cluster escapees in the Galactic halo}

\author[A Bandyopadhyay et al.]{
Avrajit Bandyopadhyay,$^{1}$\thanks{E-mail: avrajit.india@gmail.com}
Sivarani Thirupathi,$^{1}$
Timothy C Beers$^{2}$
and A.Susmitha$^{3}$
\\
$^{1}$Indian Institute of Astrophysics, Bangalore, 560034 India\\
$^{2}$Department of Physics and JINA Center for the Evolution
of the Elements, University of Notre Dame, Notre Dame, IN, 46656, USA\\
$^{3}$Tata Institute of Fundamental Research, Mumbai, 400005, India
}

\date{Accepted XXX. Received YYY; in original form ZZZ}

\pubyear{2020}

\def\LaTeX{L\kern-.36em\raise.3ex\hbox{a}\kern-.15em
    T\kern-.1667em\lower.7ex\hbox{E}\kern-.125emX}

\begin{document}

\label{firstpage}

\maketitle

\label{firstpage}
\pagerange{\pageref{firstpage}--\pageref{lastpage}}
\maketitle

\begin{abstract}
The stars SDSS~\sdsszerosix and SDSS~\sdssonenine are relatively bright stars that were initially observed as part of the SDSS/MARVELS
pre-survey. They were selected, on the basis of their weak CH $G$-bands,
along with a total of 60 others, in the range of halo globular cluster metallicities
for high-resolution spectroscopic follow-up as a part of the HESP-GOMPA
 survey (Hanle Echelle SPectrograph -- Galactic survey Of Metal Poor
 stArs).
The stars exhibit typical nucleosynthesis signatures expected from the so-called second-generation
stars of globular clusters. The light-element anti-correlation of 
Mg-Al is detected, along with elevated abundances of Na. Carbon is found to be depleted,
which is compatible with expectation. Lithium is also detected in SDSS~\sdsszerosix and SDSS~\sdssonenine; the measured abundances are similar to those of normal halo giant stars.  These bright escapees provide a unique opportunity to study the nucleosynthesis events of globular
clusters in great detail, and shed light on their chemical-enrichment histories.
\end{abstract}




\section{Introduction}

Globular clusters (GCs) are expected to lose a large amount of stellar
mass during their interactions with the tidal field of the Milky Way (e.g.,
\citealt{bm2003}; \citealt{halom}; \citealt{martellreviewlike}, and references therein).
 There are several processes by which stars can escape from GCs. The primary reason is their lower bounding energy, as they are of low mass due to the equipartition of energy, which in turn causes mass segregation. The various effects of mass segregation on the mass of escaping stars are investigated in \citet{balbig}. Among many others, \citet{bm2003} also discusses such effects on clusters embedded in tidal fields. Other processes, such as
disc shocking and dynamical friction, can also contribute to the loss of
stars from GCs. However, mass loss in Milky Way GCs due to dynamical friction would be negligible, as the mass loss in this scenario primarily depends on the Galactocentric distance of the cluster. Although disc/bulge shocks do produce enhanced mass loss (e.g., \citealt{dehnen2004}), most of the mass loss is due to secular evolution.  The importance of each mass-loss process depends upon
the properties of the cluster, as shown in the ``vital diagram" of
\citet{ostriker97}. 

Chemical tagging is one of the important tools for identifying stars of
GC origin among the halo field stars \citep{freemanhawthorn}. Although
both the cluster and field populations exhibit similar elemental abundances of
$\alpha$-, Fe-peak, and neutron-capture elements \citep{gratton2004,
pritzl2005, lind2015}, many stars in GCs exhibit certain unique trends for
their light elements (C, N, O, Na, Mg, Al), uncharacteristic of
the vast majority of halo stars (e.g., \citealt{kraftb, krafta,
norris79}, and numerous references since). These traits are thought to
emerge as a result of self pollution within the cluster, where the gas
from the first generation of stars does not escape from the cluster, but
instead pollutes the second-generation stars with the products of
advanced hydrogen burning \citep{kraft1997, carretta2009a,
carretta2009b}. In hotter regions (T $>$ 40 MK), the Ne-Na chain begins
converting $^{20}$Ne to $^{23}$Na, while simultaneously O is
depleted via the ON cycle \citep{gratton2012}. At still higher
temperatures (T $>$ 70 MK), the Mg-Al cycle is initiated, which
steadily depletes $^{24}$Mg and then $^{25}$Mg to $^{27}$Al
\citep{denissenkov96, salaris2002}. The sites where the C-N-O, Ne-Na,
and Mg-Al cycles occur is the hot-bottom burning (HBB) regions
\citep{bs91, booth92} of the outer convective envelope of intermediate-to-high mass (3-8 M$_\odot$) AGB (IH-AGB) stars. Sufficiently high temperatures are not attained in the H-burning shells of low-mass RGB stars to sustain these reactions, hence no such variations are expcted to be observed in halo field stars  \citep{grattonaanda2000}.

Fast-rotating massive stars (FRMS) are also a probable site for similar
nucleosynthesis reactions to take place \citep{decressin2007}. The winds
from these early generations of IH-AGB stars and FRMS alter the chemical
composition of the birth clouds of subsequent generations of stars.
Consequently, a large fraction of present-day GC stars are enhanced in
Na and Al, along with depleted levels of C, O, and Mg. Almost all the
Galactic GCs have been found to host multiple generations of stars
\citep{bragaglia17}, as traced by the C-N-O anomaly \citep{kayser2008,
smolinsky2011} and the anti-correlations of Na and O, and Mg and Al
(e.g., \citealt{carretta2009b, carretta2009a}; \citealt{Lee2010}; \citealt{martellg};
\citealt{martell2011};  \citealt{Bragaglia2015}; \citealt{Carretta2017}).The scenarios leading to the onset of the multiple stellar populations are highly debated; there could be several scenarios, as comprehensively discussed in \citet{bastianlardo2018}.

\citet{martellg} demonstrated that $\sim$2.5\% of $\sim$2000
low-metallicity halo giants studied by the SEGUE (Sloan Extension for
Galactic Understanding and Exploration; \citealt{yanny2009}) survey
exhibited enhanced N and depleted C relative to the remaining field
stars. A similar result was obtained by \citet{martellsegue} from
SEGUE-II. \citet{rammel2012} discovered two field dwarfs showing the
Na-O anti-correlations, which could be attributed to GC origin.
\citet{lind2015} found one star from the GAIA-ESO survey with Mg and Al
abundances largely different from the halo population, but consistent
with GC abundances. Halo giants with GC-like abundances of N and Al were
also identified by \citet{martellapogee2016}, using the SDSS APOGEE
(Apache Point Galactic Evolution Experiment; \citealt{apogee2016})
survey.

These studies have estimated that a substantial portion of the halo could
have been contributed by GCs (e.g., \citealt{carretta2010x};
\citealt{rammel2012}; \citealt{lind2015}), but consensus has not 
yet been reached regarding the precise fraction.
\citet{martellg} and \citet{martell2011} estimated that 3\% of the halo could
have been contributed by GCs, consistent with the studies conducted by
\citet{carretta2010x} and \citet{rammel2012}.
\citet{martellapogee2016} found 2\% of their sample of halo stars
to have the chemical signatures of second-generation GC stars, which
corresponds to a much larger amount of total mass loss. Following
\citet{martellsegue}, the original contribution from GCs amounts to 13\%
of halo stars, in order to account for the 2\% being chemically taggable
as second-generation stars. From the DR14 release of SDSS-IV, 
\citet{koch2019} have further refined the observed fraction of such objects to be 2.6\%, bringing the original contribution from GCs to $\sim$11\%.

Here we discuss the discovery of two likely second-generation GC
escapees with enhanced Na, Al, and depleted C, O, and Mg.

\section{Observations and Analysis}

High-resolution ($R \sim 30,000$) spectroscopic observations of our
two program stars, \sdsszerosixf and \sdssoneninef 
(hereafter \sdsszerosix and \sdssonenine)
were carried out as a part of the GOMPA (Galactic
survey Of bright Metal Poor stArs) survey, using the Hanle Echelle
Spectrograph on the 2-m Himalayan Chandra Telescope (HCT) at the
Indian Astronomical Observatory (IAO). The targets for the HESP-GOMPA survey were selected from
the spectroscopic pre-survey of MARVELS, which
was carried out as a part of SDSS-III. This offers the
opportunity to identify bright halo stars which could be studied at high
spectral resolution using moderate-aperture telescopes. The pre-survey uses
simple magnitude and color cuts ($8 < V < 13$; $B-V > 0.6$) to select
targets suitable for the MARVELS RV survey. The survey fields are mostly
low-latitude fields, which is suitable for exo-planet searches, but
not ideal for detecting metal-poor stars. We have used synthetic
spectral fitting of the pre-survey data to identify new metal-poor
candidates (in the domain of GC metallicity; [Fe/H] $> -2.5$) with weak
CH $G$-bands, a well-known feature of GC stars that can be studied from
low-resolution data (e,g., SDSS). We have obtained high-resolution data
for 60 metal-poor stars in the metallicity range of GCs (the survey paper is in prep.).
Two stars among them were found to be likely GC escapees -- showing all
the expected chemical signatures in their high-resolution spectra. The stars were observed at a spectral resolution of $R \sim 30,000$ over the wavelength
range 380nm to 1000 nm. Details of the observations, along with the
signal-to-noise ratios and $V$ magnitudes, for these two stars are listed in Table 1.

Data reduction was carried out using the IRAF echelle package,
as well as the publicly available data
reduction pipeline for HESP developed by Arun Surya. A cross-correlation
analysis with a synthetic template spectrum was carried out to obtain the radial velocity (RV) for each star, listed in Table 1.

Photometric as well as Spectroscopic data have been used to estimate the
stellar atmospheric parameters for these stars. $T_{\rm eff}$, $\log (g)$, [Fe/H],
and microturbulent velocity.  The abundances of individual elements
present in each spectrum were determined using standard procedures,
as described in \citet{bandyopadhyay}. Photometric temperatures
were obtained using the available data in the literature and the
standard $T_{\rm eff}$-color relations derived by \citet{alonso1996} and
\citet{alonso1999}. $T_{\rm eff}$ estimates have also been derived
spectroscopically, demanding that there be no trend of Fe~I line
abundances with excitation potential, as well as by fitting the
$H_{\alpha}$ profiles. The wings of $H_{\alpha}$ are also sensitive to temperature. Estimates of surface gravity, $\log (g)$, for
these stars were determined by the usual technique that demands equality
of the iron abundances derived for the neutral (Fe~I) lines and singly
ionized (Fe~II) lines. Parallaxes from Gaia have also been employed to
derive the log (g) for individual stars. The wings of the Mg~I lines,
which are sensitive to variations of $\log (g)$, have also been
fitted to obtain the best-fit value. The plots for ionization balance
and using Fe I and Fe II lines are shown in the top panels of Figure 1;
fits for the Mg triplet and $H_{\alpha}$ are shown in the middle panels for the two program stars. The wings of the H$\beta$ line is sensitive to both variation in temperature and log(g). Spectral fits of H$\beta$ 
 line are shown in the bottom panel of Figure 1, with the best fit marked in red.
For determination of the stellar parameters, 81 and 99 Fe I
lines and 12 and 8 Fe II lines could be measured for \sdsszerosix and
\sdssonenine, respectively. The adopted
values of the stellar parameters for both stars are shown in
Table 1. 

We have employed one-dimensional LTE stellar atmospheric models (ATLAS9;
\citealt{castellikurucz}) and the spectral synthesis code TURBOSPECTRUM
\citep{alvarezplez1998} for determining the abundances of the individual
elements present in each spectrum. We have considered the equivalent
widths of the absorption lines present in the spectra that are less than
120 m{\AA}, as they are on the linear part of the curve of growth.
Version 12 of the turbospectrum code for spectrum synthesis and
abundance estimates has been used for the analysis. We have adopted the
hyperfine splitting provided by \cite{mcwilliam1998} and Solar isotopic
ratios.

We have used the method of equivalent-width analysis for the
clean, strong, and unblended lines of the light elements (such as C, N, and O), the $\alpha$-elements, and the Fe-peak elements. Spectral synthesis was
carried out for the weaker features of these elements, and all lines of the
neutron-capture elements, taking into account the hyperfine
transitions where they are present. Solar abundances for the individual elements are taken from \citet{asplund2009}.


\begin{figure}
\centering
\includegraphics[width=\columnwidth]{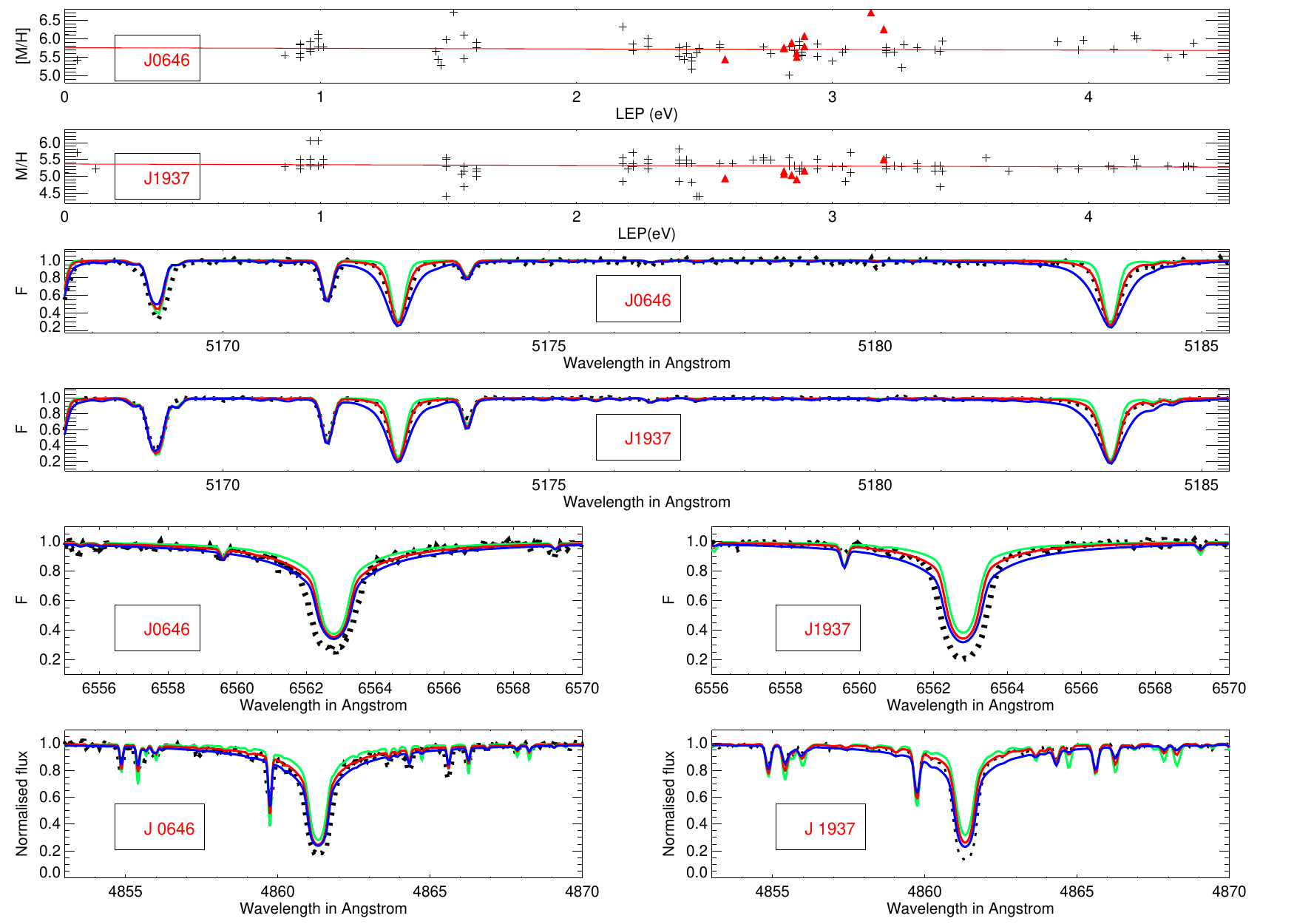}
\caption{The stellar parameters for the two program giants. The upper panels display
the ionization equilibrium plot, in which the best fit with the minimum
slope and standard deviation among all the computed model stellar
atmospheres for each star has been shown. Black crosses denote the Fe I
abundances, while red filled triangles indicate the Fe II abundances. The middle panels show
the fits in the spectral region of the Mg triplet for different values of
log(g) in steps of 0.75 dex; the best-fit value is marked in red for each star.  The lower
panels show the fits in the H-$\alpha$ region for different values of
temperature in steps of 300K; the best-fit value is marked in red for each star. The bottom panels show the fits for the wings of H$\beta$; the best fit is marked in red, and is adopted as the stellar parameter of the star. The green and blue lines show the departure variation of temperature by 200K.} \label{c4f3}
\end{figure}


\section{Abundances}

\subsection{Light and $\alpha$-Elements}

Lithium was detectable for both \sdsszerosix\ and \sdssonenine. The
strong Li doublet at $\lambda$6707\,{\AA} was used, from which we obtain 
abundances of $A$(Li) = 0.95 and 1.05, respectively, similar to 
other evolved giants observed in the field.

Carbon abundances were measured by performing a spectral synthesis for
the CH $G$-band region around $\lambda$4313\,{\AA}. The band head could be
detected and measured in \sdsszerosix\ and \sdssonenine, yielding
best-fit values of [C/Fe] = $-$0.02 and [C/Fe] = $-$0.53, respectively.
Corrections to the C abundances associated with the evolutionary state for each star \citep{placcocemp2014} have also been
computed\footnote{http://vplacco.pythonanywhere.com/}, 
and found to be +0.01 dex and +0.50 dex for \sdsszerosix\ and
\sdssonenine, respectively.

Nitrogen abundances were obtained by measuring the CN molecular band at $\lambda$3883\,{\AA}. The C abundances obtained from the $G$-band were used, with a wide range of N
abundances, and the best fit of the spectral band head was taken as the
value of the N abundance. However, being close to the extreme blue end
of the spectrum, the signal-to-noise in this region is poor, and thus only an upper limit could be derived for one of our stars, \sdssonenine; it is found to be slightly enhanced, with a value of [N/Fe] $<$ +0.42. Oxygen abundances were measured from the weak lines at $\lambda$6300\,{\AA} and
$\lambda$6363\,{\AA}, which are regions that are heavily populated by atmospheric
lines, thus telluric correction is vital for obtaining the correct
values of oxygen. O abundances could be measured for \sdssonenine while an upper limit could be obtained for \sdsszerosix;
both are found to be enhanced, with abundances of [O/Fe] $=$ +0.31 and
[O/Fe] $<$ +0.51, respectively.  

Among the $\alpha$-elements, Mg and Ca could be measured for both the
stars, while the Si lines were too weak to derive any meaningful abundances for either.
Several lines for Mg and Ca could be obtained throughout the spectra, of
which only the clean lines were used to derive the abundances. The very
strong lines, such as the Mg triplet around $\lambda$5172\,{\AA},
were ignored in the computation of the abundances. \sdsszerosix\
is found to have [Ca/Fe] = +0.23 and [Mg/Fe] = +0.21, while
\sdssonenine\ has [Ca/Fe] = +0.21 and [Mg/Fe] = +0.30, which are
somewhat lower than the typical $\alpha$-element enhancement of +0.4 dex
in halo stars. Calcium could be taken as the true representative of the
$\alpha$-elements, as other species, such 
as O, Si, and Mg are often
altered due to the recycling of the products of an earlier generation of
stars during subsequent star formation inside GCs
\citep{kraft1997, gratton2004, carretta2010x, gratton2012}.

Na and Al are the most important among the odd-Z elements to tag a star
of GC origin, and both were detected for our stars. Aluminium abundances have been derived by spectral fitting of the strong
resonance line at $\lambda$3961\,{\AA}, while the D1 and D2 lines at $\lambda$5890\,{\AA}
and $\lambda$5896\,{\AA} were used for deriving abundances of Na. Na line at $\lambda$8194.8\,{\AA} could also be detected. The spectral
fits for Al are shown in Figure 2. The NLTE corrections for this line
could be rather high, close to +1.0 dex, as discussed by \citet{baumuller},
\citet{andrievskyal}, and \citet{nordal}. The NLTE corrections for Na \citep{andrievskyna}
have also been taken into account, and incorporated in the final values.

\subsection{Fe-Peak Elements}

The abundances of Cr, Co, Mn, Ni, and Zn could be measured by the usual
equivalent-width analysis of the clean lines. Non-LTE corrections
for each species (where available) have been incorporated in the final abundances.
Cr and Co also suffer from large NLTE corrections, which could be
a reason for the over-abundance (e.g., \citealt{bergemann}).

The two odd-Z elements Mn and Cu show the usual deficiency with respect
to Fe in the metal-poor domain, as found in previous studies of halo and
GC stars. However, Ni is found to be rather high for 
\sdssonenine. Three lines of Ni could be detected at $\lambda$4401.538\,{\AA}, $\lambda$4459.027\,{\AA}, and $\lambda$5476.920\,{\AA}; the Ni line at $\lambda$4459.027\,{\AA} yielded an anomalously high abundance. Considering the other two lines, the abundance of Ni is based on the other two lines, yielding [Ni/Fe] = +0.21 and [Ni/Fe] = +0.28 for \sdsszerosix\ and \sdssonenine, respectively. The discrepancy in the Ni abundance due to the spectral line at $\lambda$4459.027\,{\AA} requires further investigation.

The identical trends for Fe-peak elements in GCs and halo field stars indicate that they
likely had a similar pre-enrichment history during the epoch of formation in
the early phases of Galactic chemical evolution.

\subsection{Neutron-Capture Elements}

The neutron-capture elements Sr and Ba could be measured for both the program stars, and found to have normal abundances without notable
irregularities. The resonance lines of Sr II at $\lambda$4077\,{\AA} and $\lambda$4215\,{\AA} were used to derive the Sr abundances, while the resonance lines
at $\lambda$4554\,{\AA} and $\lambda$4934\,{\AA} were measured to determine the Ba abundances. 
Spectral synthesis was employed to derive both abundances, taking into 
account the hyperfine transitions. Spectral fits for Ba are shown in Figure 2 for both program stars.

The complete abundance tables for \sdsszerosix and \sdssonenine are provided in Tables 2 and 3 respectively.

\begin{figure}
\centering
\includegraphics[width=\columnwidth]{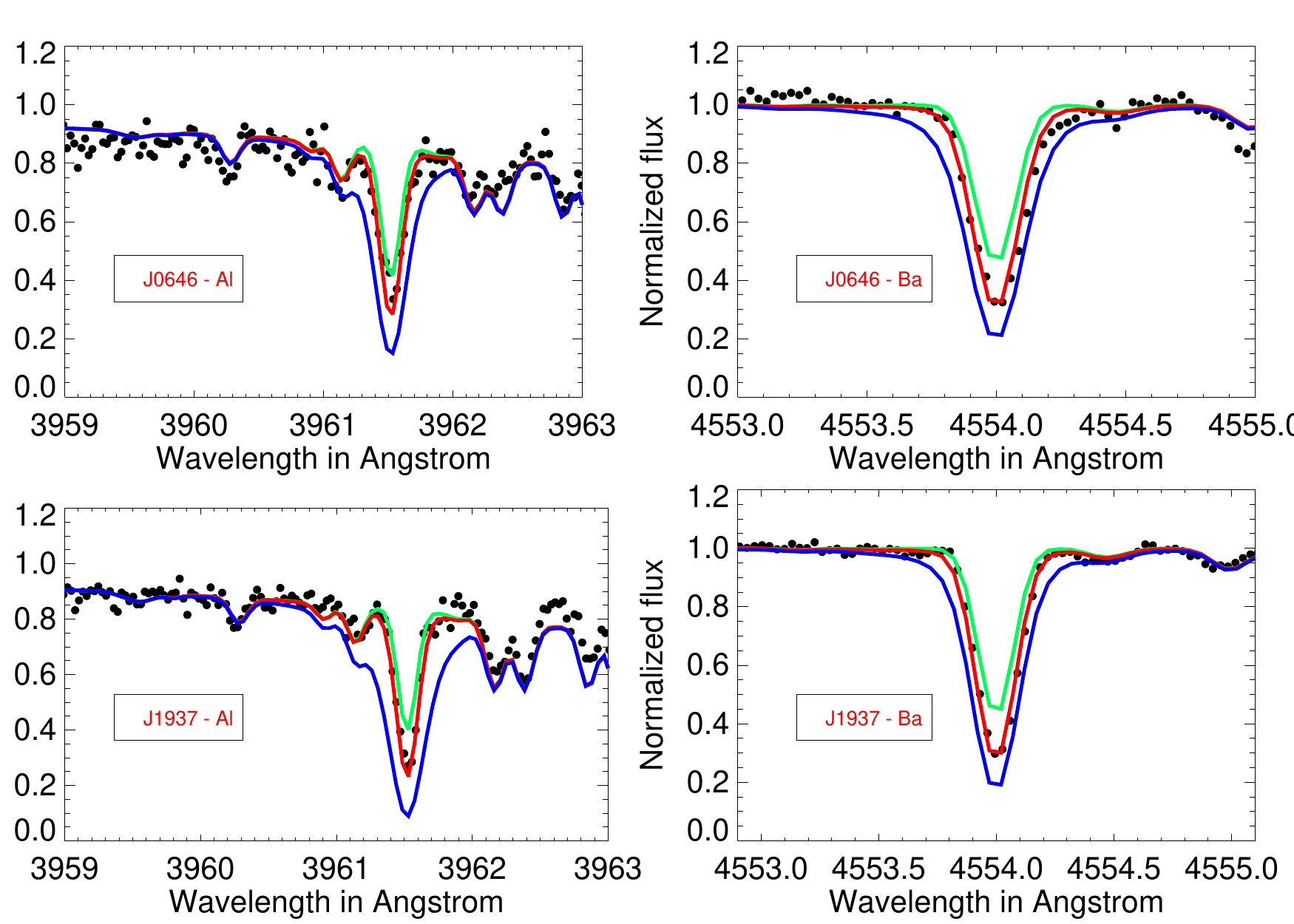}
\caption{The left
panels show the fits for Al, while the right panels show the fits for
Ba. The red lines denote the best-fit abundances, over-plotted with two synthetic spectra 
of abundances 0.25 dex higher and lower in blue and green, respectively. The names of the program stars and corresponding synthesized elements are provided in each panel.}
\end{figure}

\begin{figure}
\centering
\includegraphics[width=\columnwidth]{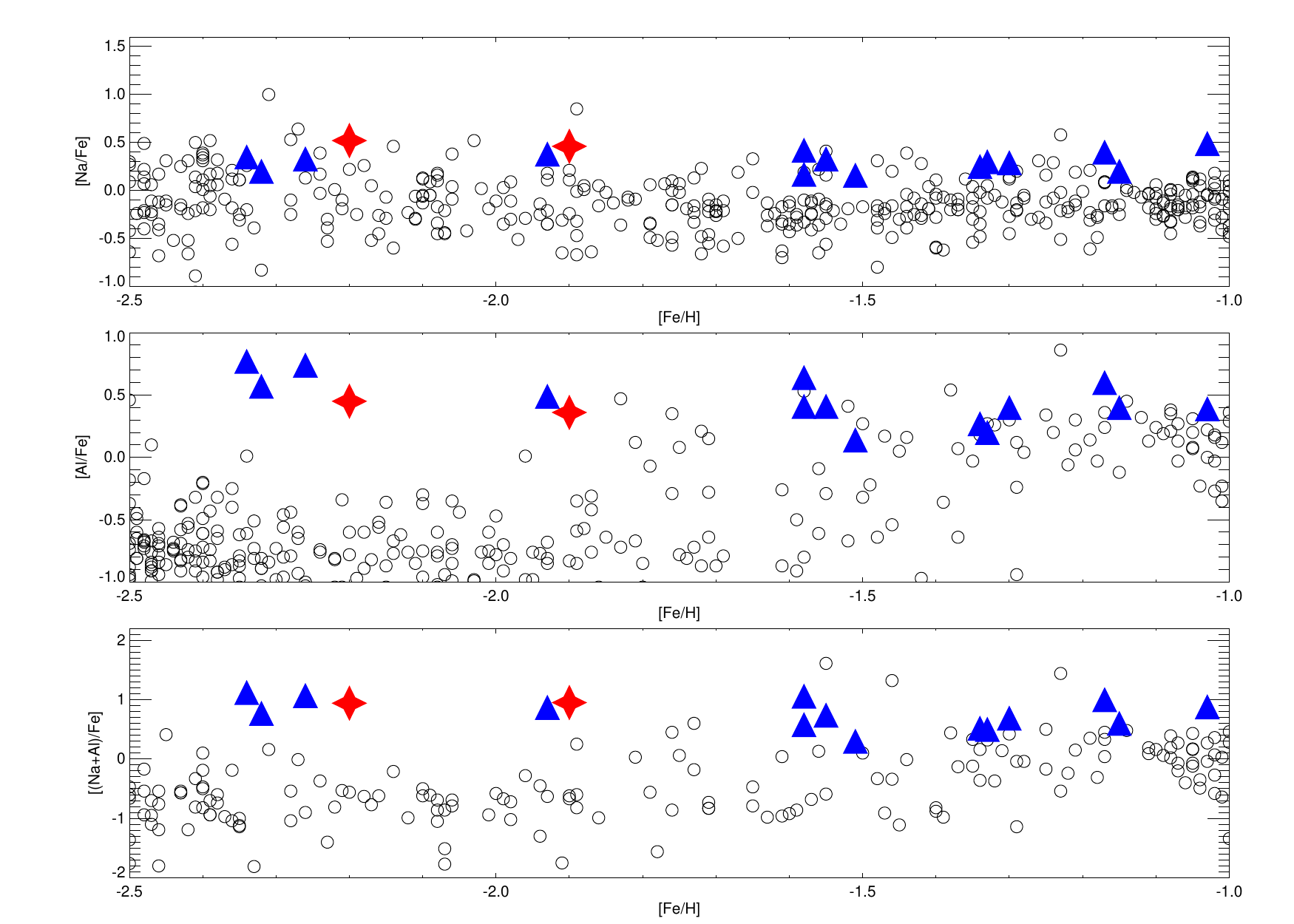}
\caption{A comparative study on the light-element abundances for GCs and
halo stars. The blue upward triangles indicate mean GC abundances,
whereas the open circles indicate the halo stars. The two program
stars discussed here are marked with red filled stars. The three panels
show the distribution of the key elements Na and Al with metallicity. Our program
stars consistently fall in the domain of GC abundances in all the plots. The data
for GCs are taken from Carretta et al. (2009a), and the abundances for
halo stars are taken from the SAGA database (Suda et al. 2008).}
\end{figure}

\begin{figure}
\centering
\includegraphics[width=\columnwidth]{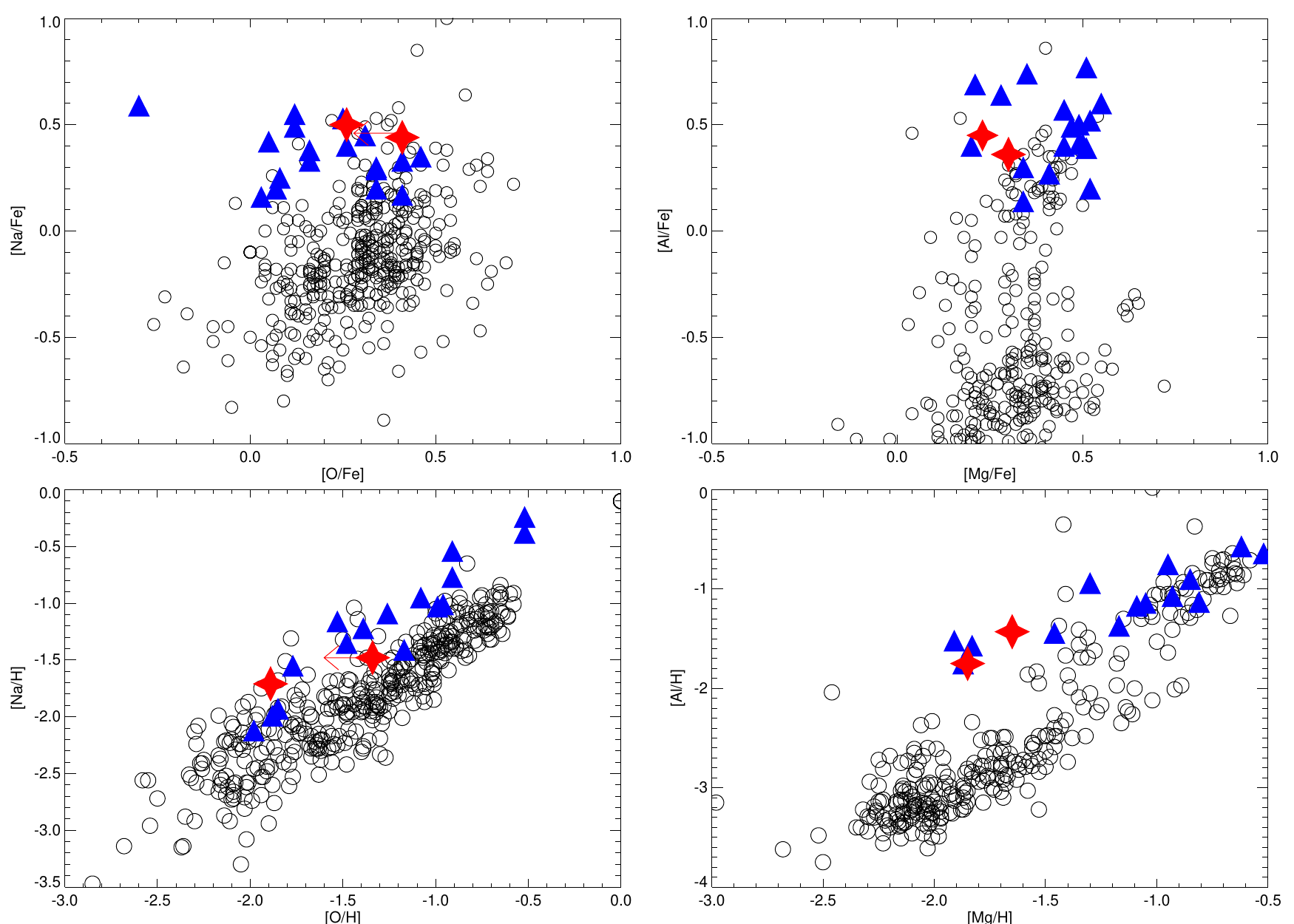}
\caption{The top two panels
show the Na-O and Mg-Al anti-correlations. 
The two panels at the bottom probe the origin of light-element
anti-correlations by removing the dependence on metallicity. The red arrow indicates the obtained upper limit for O abundances in
the case of \sdsszerosix.The data
for GCs are taken from Carretta et al. (2009b), and the abundances for
halo stars are taken from the SAGA database (Suda et al. 2008). The symbols are the same as in Figure 3.}
\end{figure}

\begin{figure}
\centering
\includegraphics[width=\columnwidth]{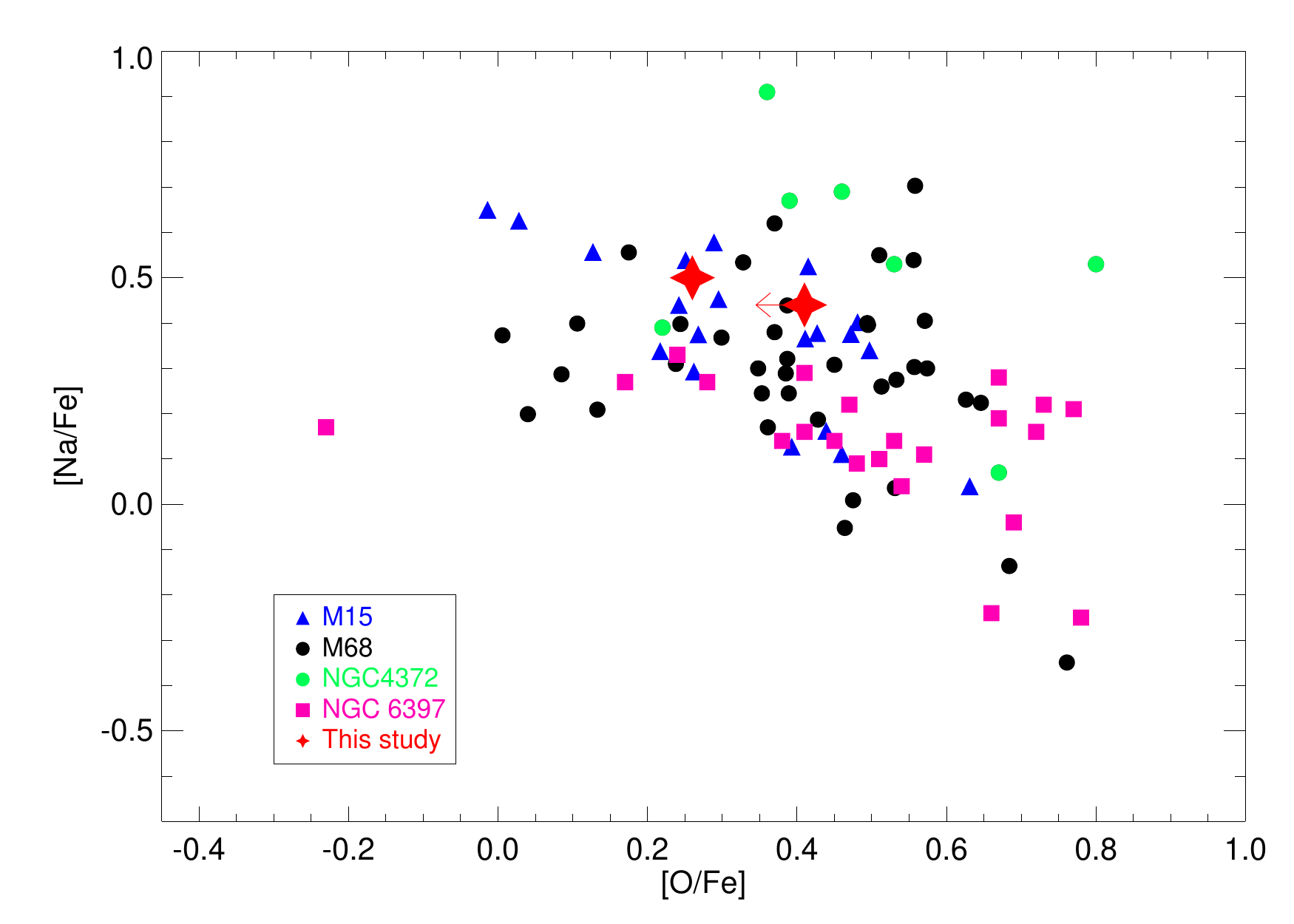}
\caption{\small Comparison of the abundances of the two halo field giants with the other metal poor globular cluster in the same range of 
metallicity namely M15, M68, NGC 4372 AND NGC 6397 along the Na-O anticorrelation. The abundances of the individual stars in these 
GCs have been compiled as follows - M15 were obtained from Carretta et al. (2009b) ; M68 were taken from Lee  et  al.  (2004)
and Carretta et al. (2009a); NGC 4372 were obtained from  San Roman, I. et al. (2015); NGC 6397 were taken from Carretta et al. (2009b), 
 Lind, K.et al. (2011) and Pasquini, L. et al. (2008).}
\end{figure}

\section{Results and Discussion}


Both of our program stars exhibit enhancement in [Na/Fe] and [Al/Fe]
compared to halo stars of similar metallicity. They also show an
under-abundance of [C/Fe], along with depletion in [O/Fe], which is
compatible with second-generation GC stars. [Mg/Fe] is also found to be
depleted in \sdsszerosix\ and \sdssonenine. Since elements such as C, O, and Mg could be
significantly altered during quiescent burning in proton-fusion
reactions \citep{gratton2004}, Ca should be adopted as the best
representative of the $\alpha$-elements for comparison with halo-star
abundances. The degree of $\alpha$-element enhancement based on the Ca abundances is found to be slightly lower than is typical for 
halo stars, at [$\alpha$/Fe]= +0.22.

We have conducted a comparative study of the light-element abundances
for our program stars with stars from the halo population and GC
abundances, as shown in Figure 3. The abundances for the GC population are based on the UVES spectra of 19 Galactic GCs, as reported 
by \citet{carretta2009b} while the abundances for
halo stars are taken from the SAGA database \citep{sudasaga}.
The top and middle panels display the enhancement in [Na/Fe] and [Al/Fe], with
respect to [Fe/H], for GC stars and halo stars. To more clearly separate
the GC population from the halo stars, we have also plotted [(Na+Al)/Fe]
vs. [Fe/H] in the bottom panel. Both the program stars fall in the domain of GC
abundances in these plots.

We note that the observed enhancement in [Na/Fe] and [Al/Fe] could also arise if these stars were members of the thick disc, but 
such stars are expected to show a much higher metallicity than our program stars, and are also inconsistent with the C and O 
abundances. The space velocities have also been determined for the target stars, and are found to be consistent with space 
velocities for halo stars \citep{kinman2007}. The values of u,v and w for \sdssonenine are  -165.4, -172.9, and -92.3 while 
for \sdsszerosix they are found to be  330.3 -275.3 and -57.0. Figure 4 in \citet{kinman2007} shows the distribution of stars, 
based on their space velocities, and classifies them  into prograde and retrograde orbits. \sdssonenine is consistent with the 
halo stars with prograde orbits, while \sdsszerosix falls very close to the edge of the line separating prograde motion from 
retrograde motion. Both of the program stars are on halo-like orbits. Thus, thick-disc membership appears improbable from the 
consideration of both abundances and space velocities.

To strengthen the argument, we have also tried to look for the Na-O and
Mg-Al anti-correlations in Figure 4. The data
for GCs are taken from \citet{carretta2009b}, and the abundances for
halo stars are taken from the SAGA database \citep{sudasaga}. Though we could only obtain an upper limit for O in \sdsszerosix the abundances fit better 
with the GC  population. 
In the bottom panels of Figure 4, we attempt to probe the existence of
these anti-correlations by removing the trends with metallicity. In the
[Al/H] vs. [Mg/H] plane, the GCs and the likely escapees still stand out
from the halo population, and exhibit a different trend in the
distribution, whereas an offset could be seen with a similar trend
between the GCs and halo population in the [Na/H] vs. [O/H] plane. 

Figure 5 provides a comparison of the abundances of our two program stars with individual GC stars of similar metallicity from M15,
M68, NGC 4372 and NGC 6397; the Na-O anticorrelation is clear. The abundances of the individual stars in these 
GCs have been compiled as follows - M15 were obtained from \citet{carretta2009b} ; M68 were taken from \citet{lee2004} 
and \cite{carretta2009b}; NGC 4372 were obtained from \citet{sanroman}; NGC 6397 were taken from \citet{carretta2009b}, 
\cite{lindcharbonnel2010} and \citet{pasquini2008}. The position of the target stars in the Na-O plane is found to be 
well within the scatter of the GC stars of similar metallicity.

Lithium could also be detected in both the program stars, and is found to be normal. Li is a fragile element, which is completely 
destroyed in a temperature range much 
lower than that required for operation of Mg-Al cycle. Thus, the presence of Li in second-generation stars indicates heavy dilution 
of the gas processed by p-capture reactions 
with unprocessed gas that still preserves the standard Population II lithium 
abundance \citep{dantona2019}. \\

In the case of AGB stars, Li 
could also be produced by \citet{cameronfowler1971} mechanism in the envelope of the 
star at the early stages of the HBB phase (mass loss during the Li-rich phase of the
 AGB is discussed in detail in \citet{ventura2005}). 
\citet{dantona2019} shows the classic dilution scheme for the
 abundance of Li in second-generation GC stars using AGB ejecta and mixing with 
primordial gas. As observed by \citet{vdorazi14} and \cite{vdorazi15}, the first generation (FG) and the second generation (SG) 
stars in M12 and NGC 362 exhibit the same Li abundance, which indicates the presence of a progenitor population like AGB stars that 
can produce Li. However, the simple dilution model fails to explain the internal variation and complex Li abundances in some of the 
GCs, like NGC 1904 and NGC 2808 -- the production efficiency of Li also depends upon the cluster's mass and metallicity
\citep{vdorazi15}. Li have been measured in several Galactic globular clusters; the Li abundances exhibit a similar distribution 
as normal halo stars.

\section{Conclusion}

A sample of $\sim$60 stars in the domain of GC metallicity with 
weak carbon molecular CH $G$-bands selected from the low-resolution 
SDSS/MARVEL pre-survey have been observed at high spectral resolution
to identify signatures of second-generation
stars in GCs. Two such stars were found to be
consistent with all of the expected light-element anomalies. 
The stars studied here are most likely to be GC escapees. Binary mass transfer from an
intermediate AGB or direct pollution from a massive star wind might be unlikely to
have caused the abundance anomaly due to the presence of lithium in both
these objects. Upcoming massive spectroscopic surveys will identify more such objects. 
GAIA kinematics and accurate ages from asteroseismology will throw light on the
origin and frequency of such objects.

\section{Acknowledgement}
We thank the referee for useful comments that helped clarify the presentation and enhance the quality of the paper. We are particularly thankful to the referee for suggesting the addition of Figure 5. We thank the staff of IAO, Hanle and CREST, Hosakote, that made these
observations possible. The facilities at IAO and CREST are operated by
the Indian Institute of Astrophysics, Bangalore. We also thank Prof. Piercarlo Bonifacio for his valuable comments and suggestions. T.C.B. acknowledges
partial support from grant PHY 14-30152 (Physics Frontier
Center/JINA-CEE), awarded by the U.S. National Science Foundation (NSF).
T.C.B. also acknowledges partial support from the Leverhulme Trust (UK),
which hosted his visiting professorship at the University of Hull during
the completion of this study.

\onecolumn

\begin{table}
\begin{center}
\caption{Observational Details for our Program Stars}
\begin{tabular}{|cccccccccccc|}
\hline\hline
Object &RA &DEC &Exposure &Frames &SNR &V      &Radial Vel.    &$T_{\rm eff}$ & log(g) & $\xi$ &[Fe$/$H]\\
 & & & (secs)  &       &    & (mag) & (km~s$^{-1}$ &    (K)      & (cgs)  & \\
\hline
SDSS J064655.6+411620.5 &06 46 55.6 &+41 16 20.5 &2400 &6 &43 &11.14 &$-$285.0 &5150 &2.25 &1.50 &$-$1.90\\
SDSS J193712.0+502455.5 &19 37 12.01 &+50 24 55.50 &2400 &3 &130 &10.44 &$-$184.0 &4800 &1.50 &1.50 &$-$2.20\\
\hline
\end{tabular}
\end{center}
\end{table}

\begin{table}
\begin{center}
\caption{Elemental Abundances for SDSS J064655.6+411620.5} 
\begin{tabular}{|cccccccc|}
\hline\hline
Name  &Species &Solar &lines &A(X) &[X/H] & [X/Fe] &$\sigma^*$ \\
\hline
Li &Li I &\dots   &1 &0.95 &\dots &\dots   &0.03    \\
C &CH    &8.43   &\dots &6.25    &$-$2.18 &$-$0.02   &0.05\\
O &O I    &8.69   &1 &7.30       &$-$1.39 &$+$0.51   &0.09  \\
Na &Na I &6.24 &2 &4.75        &$-$1.49 &$+$0.41 &0.08 \\
Mg &Mg I &7.60   &5  &5.91       &$-$1.69 &$+$0.21   &0.06 \\
Al &Al I &6.45   &1 &4.92        &$-$1.53 &$+$0.37   &0.11  \\
Ca &Ca I &6.34   &10 &4.67         &$-$1.67 &$+$0.23   &0.08  \\
Sc &Sc II &3.15   &3 &1.72       &$-$1.43 &$+$0.47   &0.04 \\
Ti &Ti I &4.95   &8 &3.47         &$-$-1.47 &$+$0.43   &0.05  \\
   &Ti II &4.95   &14 &3.57       &$-$1.38 &$+$0.52   &0.07\\
Cr &Cr I &5.64   &5 &3.80        &$-$1.84 &$+$0.06   &0.09 \\
   &Cr II &5.64   &3 &4.06       &$-$1.58 &$+$0.32   &0.07 \\
Mn &Mn I &5.43   &3 &3.24        &$-$2.19 &$-$0.29   &0.12 \\
Co &Co I &4.89   &2 &2.62        &$-$2.27 &$-$0.37   &0.08 \\
Ni &Ni I &6.22   &3 &4.56        &$-$1.66 &$+$0.24   &0.07  \\
Zn &Zn I &4.56   &2 &2.89        &$-$1.67 &$+$0.23   &0.06  \\
Sr &Sr II &2.87   &2 &1.25       &$-$1.62 &$+$0.28   &0.05  \\
Ba &Ba II &2.18   &3 &0.75       &$-$1.43 &$+$0.47   &0.03  \\

\hline
\end{tabular}
\end{center}
\end{table}

\begin{table}
\begin{center}
\caption{Elemental Abundances for SDSS J193712.01+502455.5} 
\begin{tabular}{|cccccccc|}
\hline\hline
Name  &Species &Solar &lines &A(X) &[X/H] & [X/Fe] &$\sigma^*$ \\
\hline
Li &Li I &\dots &1 &1.05 &\dots &\dots   &0.02  \\
C &CH    &8.43     &\dots &6.00     &$-$2.43 &$-$0.53   &0.03  \\
N &CN    &7.83  &\dots &6.05    &$-$1.78 &$+$0.42   &0.15    \\
O &O I    &8.69    &1 &6.80  &$-$1.89 &$+$0.31   &0.07   \\
Na &Na I &6.24  &2  &4.50        &$-$1.74 &$+$0.46 &0.03  \\
Mg &Mg I &7.60   &5 &5.80         &$-$1.80 &$+$0.30  &0.02 \\
Al &Al I &6.45   &1  &4.70        &$-$1.75 &$+$0.45  &0.07  \\
Ca &Ca I &6.34  &7 &4.35        &$-$1.99 &$+$0.21   &0.07   \\
Sc &Sc II &3.15   &3  &0.95        &$-$2.20 &$+$0.00   &0.02  \\
Ti &Ti I &4.95   &7 &3.14       &$-$1.81 &$+$0.39   &0.04  \\
   &Ti II &4.95  &14  &2.86        &$-$2.09 &$+$0.11   &0.04  \\
Cr &Cr I &5.64   &5 &3.33        &$-$2.31 &$-$0.11   &0.04  \\
   &Cr II &5.64   &4 &3.35        &$-$2.29 &$-$0.09   &0.06   \\
Mn &Mn I &5.43    &3 &2.77     &$-$2.66 &$-$0.46   &0.12 \\
Co &Co I &4.89   &2  &2.68     &$-$2.31 &$-$0.11   &0.03  \\
Ni &Ni I &6.22   &2 &4.70        &$-$1.52 &$+$0.28   &0.04 \\
Zn &Zn I &4.56  &2 &2.37         &$-$2.19 &$+$0.01   &0.02 \\
Sr &Sr II &2.87  &2 &0.75        &$-$2.12 &$+$0.08   &0.02  \\
Ba &Ba II &2.18   &3 &0.25         &$-$1.93 &$+$0.27   &0.02 \\

\hline
\end{tabular}
\end{center}

\end{table}

\twocolumn

\bibliographystyle{mn2e}

\label{lastpage}

\end{document}

%% file: journal_abbv.tex
\def\aj{AJ}%
\def\araa{ARA\&A}%
\def\apj{ApJ}%
\def\apjl{ApJ}%
\def\apjs{ApJS}%
\def\ao{Appl.~Opt.}%
\def\aap{A\&A}%
\def\aapr{A\&A~Rev.}%
\def\aaps{A\&AS}%
\def\mnras{MNRAS}%
\def\pasj{PASJ}%
